\documentclass[12pt]{article}
\usepackage{graphicx}
\usepackage{longtable}

\begin{document}

\makeatletter
\renewcommand*{\@cite}[2]{{#2}}
\renewcommand*{\@biblabel}[1]{#1.\hfill}
\makeatother

\title{Some Properties of Dust Outside the Galactic Disk}
\author{G.~A.~Gontcharov\thanks{E-mail: georgegontcharov@yahoo.com}}

\maketitle

Pulkovo Astronomical Observatory, Russian Academy of Sciences, Pul\-kov\-skoe sh. 65, St. Petersburg, 196140 Russia


The joint use of accurate near- and mid-infrared photometry from the 2MASS and WISE
catalogues has allowed the variations of the extinction law and the dust grain size distribution in high
Galactic latitudes ($|b|>50^{\circ}$) at distances up to 3 kpc from the Galactic midplane to be analyzed. The
modified method of extrapolation of the extinction law applied to clump giants has turned out to be efficient
for separating the spatial variations of the sample composition, metallicity, reddening, and properties of the
medium. The detected spatial variations of the coefficients $E_{(H-W1)}/E_{(H-Ks)}$, $E_{(H-W2)}/E_{(H-Ks)}$,
and $E_{(H-W3)}/E_{(H-Ks)}$ are similar for all high latitudes and depend only on the distance
from the Galactic midplane. The ratio of short-wavelength extinction to long-wavelength one everywhere
outside the Galactic disk has been found to be smaller than that in the disk and, accordingly, the mean dust
grain size is larger, while the grain size distribution in the range $0.5-11$ microns is shifted toward coarse dust.
Specifically, the mean grain size initially increases sharply with distance from the Galactic midplane, then
decreases gradually, approaching a value typical of the disk at $|Z|\approx2.4$ kpc, and, further out, stabilizes
or may increase again. The coefficients under consideration change with coordinate $Z$ with a period of
about $1312\pm40$ pc, coinciding every $656\pm20$ pc to the south and the north and showing a significant
anticorrelation between their values in the southern and northern hemispheres at intermediate $Z$. Thus,
there exists a unified large-scale periodic structure of the interstellar medium at high latitudes within at
least 5 kpc. The same periodic variations have also been found for the extinction coefficient $R_V$ within
600 pc of the Galactic midplane through the reduction of different photometric data for stars of different
classes.

\newpage
\section*{INTRODUCTION}

The wavelength dependence of interstellar extinction
(extinction law) is related primarily to the dust
grain size distribution in the absorbing medium. In
the most widespread media, the extinction maximum
in a region of space occurs at a wavelength that
coincides in order of magnitude with the mean dust
grain diameter in this region (Bochkarev 2010). For
example, the commonly considered extinction $A_V$ in
the photometric $V$ band with a wavelength of 0.55 microns
and, in general, the visual extinction is produced predominantly
by submicron sized dust grains, whereas
the infrared (IR) one is produced by supermicron
ones. Conclusions about the physical properties of
dust in the region of space under consideration and
about the contribution of dust to the total mass of
the matter can be reached by studying the wavelength
dependence of extinction. Conversely, the extinction
can be calculated by studying the properties of the
dust medium.

Numerous studies of the dust and extinction
in the Galactic disk, including those in the solar
neighborhood, have shown that submicron dust
grains producing an essentially selective extinction,
i.e., one that decreases appreciably with increasing
wavelength in such a way that the IR extinction is
insignificant compared to the visual one, dominate
here.

The dust and extinction in regions of the Galaxy
far from the Sun at its high latitudes outside the disk
have been studied more poorly. Clearly, the extinction
here is low. However, the ratio of visual to IR extinction
and, accordingly, the content of large dust grains
in the interstellar medium is still difficult to estimate
here.

IR photometry of stars in a small region of space
or a small number of stars is useless for these estimations,
because some of the stars of various classes
exhibit an unpredictable IR emission. Only averaging
the data for many stars in regions of space that definitely
exceed the typical size of an interstellar cloud
($>10$ pc) allows its influence to be smoothed out.

Only the accurate multicolor broadband IR photometry
for millions of stars over the entire sky from
the 2MASS and WISE catalogues obtained in recent
years allows the coarse dust and IR extinction outside
the Galactic disk to be investigated for the first time.
The 2MASS (2-Micron All-Sky Survey) catalogue
(Skrutskie et al. 2006) was produced in 2006 as a result
of ground-based observations in 1997-2001 and
contains photometry for more than 400 million stars
over the entire sky in the near-IR $J$ (1.23 microns), $H$
(1.64 microns), and $Ks$ (2.15 microns) bands. The WISE catalogue
(Wright et al. 2010) was produced in 2012 as a
result of the observations performed in 2010 with the
Wide-field Infrared Survey Explorer space telescope
and contains photometry for more than 500 million
stars over the entire sky, including almost all 2MASS
stars, in the mid-IR $W1$ (3.3 microns), $W2$ (4.6 microns), $W3$
(11 microns), and $W4$ (22 microns) bands.

IR surveys of selected sky regions, for example,
the mid-IR survey of the region toward the Galactic
center with the IRAC camera of the Spitzer space
telescope, have also been made in recent years. The
data from this survey used together with the 2MASS
data allowed Zasowski et al. (2009) to detect largescale
systematic spatial variations of the extinction
law and, consequently, the sizes and other properties
of dust grains in the inner (relative to the Sun) Galactic
disk within 15 kpc. Supermicron- and submicron-sized
dust apparently dominates in the parts of the
disk close to the Galactic center and in the solar
neighborhood, respectively. These variations have
such a large scale that they are actually characteristic
of a diffuse medium rather than small dense clouds
and star-forming regions.

Gontcharov (2013) tested the results by Zasowski
et al. (2009) in the first and fourth Galactic quadrants
of the disk using WISE data instead of Spitzer ones
and carried out a similar search for the spatial variations
of dust properties in the second and third Galactic
quadrants, where only the 2MASS and WISE data
are available. The result by Zasowski et al. (2009)
was confirmed for the inner (relative to the Sun)
Galactic disk: the mean dust grain size decreases
with increasing Galactocentric distance. However,
in the outer Galactic disk that was not considered by
Zasowski et al. (2009), this trend is reversed: the dust
grains at the disk edge are, on average, larger than
those in the solar neighborhood. These variations
refer to the region with a radius of more than 5 kpc
around the Sun.

When analyzing the extinction law and dust
properties in certain directions at latitudes $|b|=10^{\circ}$,
Gontcharov (2013) found the mean dust grain size to
increase with distance from the Galactic midplane.

This confirms the previous result by Gontcharov
(2012a) presented in the three-dimensional map of
variations in extinction coefficient $R_V\equiv A_V/E_{(B-V)}$
within 600 pc of the Sun. This map was obtained
using visual photometry from the Tycho-2 catalogue
(H\o g et al. 2000) and 2MASS IR photometry.
According to this map, $R_V$ changes systematically
with distance from the Galactic midplane. It has
a sharp minimum in the layer with a thickness of
less than 200 pc near the midplane, increases with
distance from the midplane, reaching its maximum
at distances $|Z|\approx150\div400$ pc from the midplane,
and slightly decreases at $|Z|\approx500$ pc. Figure 1
shows the detected variations of $R_V$ with heliocentric
distance for the region with a radius of $8^{\circ}$ around
the Galactic north (black diamonds, curve, and vertical
error bars) and south (grey squares, curve, and
vertical error bars) poles. This figure can be compared
with Fig. 6 from Gontcharov (2012a). The
coefficient $R_V$ is seen to have a classical value of 3.1
near the Sun. In addition, the changes of $R_V$ to the
south and the north anticorrelate at $|Z|>100$ pc: the
increase to the north of the equator corresponds to the
decrease at the same distance to the south and vice
versa. This anticorrelation is discussed below.

The large-scale general Galactic trends in the
variations of dust properties can be explained by the
influence of the spiral pattern as probably the only
place where the dust is fragmented and sorted by size.
As a result, as Gontcharov (2013) assumed, fine dust
exists only in the part of the Galactic disk where the
spiral pattern operates, i.e., far from both the Galactic
center and the edge. The fraction of fine dust at the
Galactic center, at the disk edge, and outside the disk
should then be minimal. This study is devoted to testing
this assumption regarding the Galactic regions
outside the disk. For the convenience of comparing
the results, the applied method and the quantities
considered here are analogous to those in Zasowski
et al. (2009) and Gontcharov (2013).

\section*{DATA REDUCTION}

For this study, we selected stars at high latitudes
($|b|>50^{\circ}$) common to 2MASS and WISE with an
accuracy of their photometry in the $J$, $H$, $Ks$, $W1$,
and $W2$ bands better than $0.05^m$. In this case, the
sample remains almost complete in the range of magnitudes
$6^m<Ks<14^m$.

Because of the unexpected loss of its coolant
ahead of schedule, the WISE spacecraft reduced the
observing period and did not achieve a high photometric
accuracy in the $W3$ and $W4$ bands. Therefore,
the results in these bands are barely considered here.

In this study, we have made an attempt to investigate
the properties of dust outside the disk for
the first time by using only IR photometry. This
attempt is based on the advantages of jointly using
near- (2MASS) and mid-IR (Spitzer or WISE) photometry
to which attention was first drawn by Majewski
et al. (2011). One of these advantages is the
independence of the $(H-W2)$ color index on stellar
metallicity and age. These advantages are discussed
in more detail below.

The high latitudes under consideration differ from
the disk regions considered by Zasowski et al. (2009)
and Gontcharov (2013) by a smaller number of stars,
a smaller reddening/extinction, and a larger metallicity
gradient. Allowance for these peculiarities forced
us to slightly modify the method applied by Zasowski
et al. (2009) and Gontcharov (2013). They, in
turn, adapted the classical method of determining
the spatial variations of the extinction law and dust
properties, the method of extrapolation of the extinction
law (Straizys 1977; Gontcharov 2012a), for
the infrared. This method is commonly used in its
classical form to determine the coefficient $R_V$ from
the formula $A_V/E_{(B-V)}\approx1.1E_{(V-K)}/E_{(B-V)}$.
The ratio of reddenings (color excesses) $E_{(V-K)}/E_{(B-V)}$
is a characteristic of the dust grain
size distribution in the range from about 0.5 to 2 microns.
The analogous coefficients
$E_{(H-W1)}/E_{(H-Ks)}$,
$E_{(H-W2)}/E_{(H-Ks)}$, and
$E_{(H-W3)}/E_{(H-Ks)}$
considered here following Zasowski et al. (2009)
and Gontcharov (2013) characterize the extinction
law and the dust grain size distribution in the range
from about 2 to 11 microns.

The method of extrapolation of the extinction law
requires the following in the spatial cell under consideration:
(1) a representative sample of stars with the
same unreddened spectral energy distribution; (2) the
reddening gradient within this cell should be larger
than the photometric errors; and (3) to compare the
reddenings at different wavelengths, at least one of
the color indices used should redden noticeably within
the cell, i.e., the reddening and extinction should be
sufficiently large. When using only IR photometry,
the reddening and extinction are so small that we
have to consider spatial cells with a large number of
stars and to watch carefully the balance of errors. In
most of the space under consideration, we attempt
to reveal effects at a $0.01^m$ level using photometry
with a median accuracy of $0.02^m$ in each band for an
individual star. Accordingly, the median accuracy of
each color index used for an individual star is about
$0.03^m$. Therefore, we have to consider cells with tens
and hundreds of stars. The task is complicated by the
fact that high latitudes are poor in stars.

In such a situation, averaging the data for many
stars in the cell makes sense only if their spectral
energy distribution is the same and/or predictable
with an accuracy better than $0.01^m$ depending on
the known parameters. In practice, however, this
distribution changes at least in one of the bands under
consideration by more than $0.1^m$ for neighboring
classes of stars (for example, red giant branch (RGB)
stars and clump giants) as well as with metallicity and
reddening. As it turned out, only the age may not
be considered: stars with ages older than $7\times10^9$ yr
dominate at high latitudes and the spectral energy
distribution within most of the classes at such an age
changes with age within $0.01^m$.

Thus, the method of extrapolation of the extinction
law is efficient only for a large sample of highluminosity
stars with a stable spectral energy distribution.
Gontcharov (2012a) applied it for OB and
RGB KIII stars. Consistent spatial variations of
the coefficient $R_V$ were found for so different classes.
In this study, following Zasowski et al. (2009) and
Gontcharov (2013), we applied this method to clump
giants.

Clump giants are evolved stars with nuclear
reactions in their helium cores. Their evolutionary
status and characteristics were considered in detail
by Gontcharov (2008) when analyzing a sample of
97 348 such stars from the Hipparcos (ESA 1997;
van Leeuwen 2007) and Tycho-2 catalogues mostly
within 1 kpc of the Sun. In particular, their empirical
mean absolute magnitude $\overline{M_{Ks}}=-1.52^m$ obtained
by Gontcharov (2008) by taking into account the
Hipparcos parallaxes and used below is consistent,
within $0.05^m$, with the theoretical estimate
from the Padova database of evolutionary tracks
and isochrones (http://stev.oapd.inaf.it/cmd; Marigo
et al. 2008; Bressan et al. 2012; here and below, all
theoretical results of modeling the evolution of stars
were taken from the Padova database) for the mixture
of clump giants with metallicity $\mathbf Z$ from 0.004 to 0.020
and a mass from 0.9 to 3 $M_{\odot}$ (accordingly, an age up
to $10\times10^9$ yr) at mean metallicity $\mathbf Z\approx0.014$ (i.e.,
$Fe/H\approx-0.13$, slightly lower than the solar one)
and a mass of 1.4 solar masses assumed in Tycho-2.
Encompassing a larger space, the sample of clump
giants considered here should have a wide range of
metallicities, masses, and ages than the sample from
Tycho-2. However, even the variations in age up to
$14\times10^9$ yr and in $\mathbf Z$ from 0.001 to 0.020, according
to the Padova database, lead to variations in $\overline{M_{Ks}}$
within $\pm0.1^m$, an insignificant dependence of $\overline{M_{Ks}}$ 
on color indices, and a small scatter of individual
$M_{Ks}$ around $\overline{M_{Ks}}$ ($\sigma(M_{Ks})=0.3^m$). This allows the
distance $r$ for each selected star to be calculated with
an accuracy of 15\%:
\begin{equation}
\label{rph}
r=10^{(Ks+1.52+5)/5}.
\end{equation}
The extinction $A_{Ks}$ should have entered into this
formula. However, its value and uncertainty at the
latitudes under consideration ($|b|>50^{\circ}$)
are small compared to the scatter of individual $M_{Ks}$: according
to Gontcharov (2012b), $A_V<0.3^m$; then,
$A_{Ks}<0.03^m$, given the most commonly used estimates
of the extinction law, for example, from
Cardelli et al. (1989), Draine (2003), and Indebetouw
et al. (2005). Therefore, the extinction may be
neglected in this formula for high latitudes.

The IR photometry is inaccurate not only for faint
stars but also for the brightest ones ($Ks<6^m$).
Therefore, in the sample there are almost no clump
giants within 400 pc of the Sun. However, the sample
is almost complete in the range of distances from 0.4
to 3 kpc. All of the results presented below refer to
this range of distances.

The selection of clump giants as all stars in one
of the regions of enhanced star density on the $(J-Ks)-Ks$
diagram has been used many times, for
example, by Lopez-Corredoira et al. (2002), Indebetouw
et al. (2005), and Gontcharov (2013).

Figure 2 gives an example of the $(J-Ks)-Ks$ diagram
for two of the sky regions under consideration:
(a) with a radius of $8^{\circ}$ around the Galactic north pole
and (b) in the sky sector $-15^{\circ}<l<15^{\circ}$, $-61^{\circ}<b<-53^{\circ}$,
i.e., the regions, respectively, with the
smallest and largest extinction among all those considered
in this study. It can be seen from the figure
that the reddening effect here is insignificant and the
distributions of stars on the diagram in these regions
barely differ. The left, central, and right clouds of stars
consist mostly of O-F main-sequence stars, clump
giants, and red dwarfs, respectively. The admixtures
of subdwarfs, subgiants, supergiants, RGB stars, and
other classes are small compared to the main categories.
It can be seen that these three categories of
stars barely mix at $Ks<11^m$. This value has become
one of the criteria for selecting a comparatively pure
sample of clump giants, while the main criterion is
\begin{equation}
\label{sel}
0.5^m<(J-Ks)<0.76^m.
\end{equation}

The selected stars are marked by the larger signs
in the upper parts of the graphs. This criterion efficiently
separates the clump giants from the stars
with a different spectral energy distribution, which,
among other things, manifests itself in different absolute
magnitudes $M_J$ and $M_{Ks}$. In this way, not only
a high efficiency of the method but also a sufficient
accuracy of the photometric distances obtained below
from $M_{Ks}$ is achieved. In addition to other stars,
this criterion also rejects the overwhelming majority
of subgiants and RGB stars with $(J-Ks)<0.5^m$ and $(J-Ks)>0.76^m$,
respectively (in the figure,
leftward and rightward of the cloud of selected stars,
respectively). However, even on the Hertzsprung-
Russell diagram the red giant branch and the clump of
giants are mixed. Therefore, of course, the subgiants
and RGB stars cannot be completely excluded from
our sample. It is only important that their admixture
does not distort the average spectral energy distribution
for the stars in each spatial cell under consideration.
The theoretical data on the evolution of stars
from the Padova database show that the admixture of
subgiants and RGB stars for the range (2) at $Ks>7^m$
accounts for no more than 30\% of the sample
and introduces distortions into the average spectral
energy distribution for the subsamples of 200 stars
considered below by no more than $0.01^m$ in each band
under consideration. At $Ks<7^m$ the influence of the
admixture of RGB stars should theoretically be larger,
because their luminosity is, on average, higher than
that of clump giants. Below we detected this influence
when analyzing the $Ks$ -- $(H-W2)$ diagrams at
$Ks<7^m$, i.e., according to Eq. (1), at $r<500$ pc.
The results of this study refer to $r>500$ pc and,
therefore, are not affected by the admixture of bright
RGB stars.

For the range (2), as Gontcharov (2013) showed,
the admixture of dwarfs outside the disk is more
than 10\%. However, contrary to the assumption
made by Gontcharov (2013), this admixture can be
revealed by using not only the reduced proper motions
but also the $(H-W2)$ and $(Ks-W2)$ color indices,
as was proposed by Majewski et al. (2011) and is
realized below.

According to the evolutionary tracks and isochrones
of stars from the Padova database, all the
color indices of the clump giants considered here
are almost independent of their age in the range
$7\times10^9-14\times10^9$ yr but depend significantly on
their metallicity. Therefore, in contrast to Zasowski
et al. (2009) and Gontcharov (2013), who selected
the clump giants in the disk, where the metallicity
gradient is small, here we should apply a more
sophisticated selection procedure.

Figure 3 shows the theoretical spectral energy
distribution for a typical clump giant with an age of
$10\times10^9$ yr for metallicities
$\mathbf Z=0.015$ (black squares and solid curve),
$\mathbf Z=0.006$ (grey diamonds and solid curve), and
$\mathbf Z=0.004$ (black circles and dashed line).
in the (from left to right) $J$, $H$, $Ks$, $W1$, $W2$, $W3$,
and $W4$ bands expressed in magnitudes relative to
the $W2$ magnitude. These metallicities correspond to
$Fe/H\approx-0.1$ ,$-0.5$,$-0.7$, i.e., according to present
views, they reflect the transition from the nearly solar
metallicity of the thin disk to a metallicity typical of
the thick disk-halo boundary. It can be seen that,
as has been pointed out above, for such a wide range
of metallicities the $(H-W2)$ color index changes by
less than $0.01^m$. Therefore, the systematic change in
$(H-W2)$ should be attributed only to the change in
sample composition and reddening.

The $(H-Ks)$ and $(Ks-W2)$ color indices also
change only slightly with metallicity.

It should also be noted that this spectral energy
distribution corresponds to blackbody radiation in the
$J$, $H$, and $Ks$ bands, slightly deviates from it in the
$W1$ and $W2$ bands, and deviates significantly due to
the emission in the $W3$ and $W4$ bands.

It is also important that the $(J-H)$ color index
decreases with decreasing metallicity but is known
to increase with increasing extinction. This allows
the metallicity and extinction variations with distance
from the Galactic midplane (with increasing $|Z|$) to
be separated.

Figure 4 shows the theoretical isochrones before
the stage of a clump giant (highlighted by the large
symbols) for stars with metallicities $\mathbf Z=0.008$ (black
squares and solid curve), $\mathbf Z=0.004$ (grey diamonds
and solid curve), and $\mathbf Z=0.0004$ (black circles
and dashed line) on the (a) $(H-W2)$ -- $M_{Ks}$
and (b) $(Ks-W2)$ -- $M_{Ks}$ diagrams according to the
Padova database. It can be seen that almost all of
the main-sequence, RGB, and clump giant stars fall
into the narrow range of colors
\begin{equation}
\label{hk}
0.03^{m}<(H-W2)<0.1^{m}, -0.08^{m}<(Ks-W2)<0.04^{m}.
\end{equation}
The red dwarfs and RGB-tip giants turn out to be
outside this range. Thus, applying these limitations
by taking into account the possible reddening of stars,
along with the criterion (2), we free the sample of
clump giants from the main admixtures.

However, even before the rejection of stars, the
narrow range of $(H-W2)$ and its independence of
metallicity allow the reddening near the Sun and the
influence of the admixtures of red dwarfs and RGB-tip
giants to be estimated. Figures 5a and 5b show the
$Ks$ -- $(H-W2)$ diagrams for all stars (not only the
clump giants) in the regions with a radius of $8^{\circ}$ around
the Galactic north and south poles, respectively. The
curves indicate the results of our moving averaging
of the data over 15 points. It can be seen that the
mean $\overline{(H-W2)}$ both southward and northward of
the Sun has a minimum near $Ks\approx7.8^{m}$. Judging
by Fig. 4a, the decrease in $\overline{(H-W2)}$ at $6^{m}<Ks<7.8^{m}$
implies a decrease in the fraction of RGB stars
in the sample.

It can be seen that the minimum $\overline{(H-W2)}$ are
$0.08^{m}$ and $0.10^{m}$ northward and southward of the
Sun, respectively. The dereddened color at $0.005<\mathbf Z<0.015$
for most of the main sequence and the
clump giants is $(H-W2)_0=0.052^{m}$. Then, $E_{(H-W2)}=0.028^{m}$ and
$0.048^{m}$ toward the north and
south poles, respectively, at heliocentric distances of
several hundred pc, behind most of the equatorial
absorbing layer. Thus, the reddening/extinction
to the south of the Sun is larger than that to the
north. According to the most commonly used
estimates of the extinction law, for example, from
Cardelli et al. (1989), Draine (2003), and Indebetouw
et al. (2005), $E_{(J-Ks)}\approx E_{(H-W2)}$, $E_{(B-V)}\approx 2E_{(H-W2)}$
and $A_{V}\approx 6.2E_{(H-W2)}$. The reddening estimates toward the poles
obtained here are then in good agreement with those
from Gontcharov (2010), $E_{(J-Ks)}=0.02^{m}$ and $0.03^{m}$,
to the north and the south obtained when
constructing the three-dimensional reddening map
within 1.6 kpc of the Sun and with the estimates
from Gontcharov (2012b), $E_{(B-V)}=0.06^{m}$ and
$A_{V}=0.2^{m}$, toward the Galactic poles obtained
by comparing the reddening/extinction maps from
Gontcharov (2010), Schlegel et al. (1998), and Jones
et al. (2011).

Taking into account the limitations (3), the estimates
$E_{(Ks-W2)}=0.62E_{(H-W2)}$ and $E_{(Ks-W1)}=0.56E_{(H-W2)}$
according to the mentioned
extinction laws, the possible reddening of the colors
for stars farther than 700 pc, the photometric errors,
and the change in dereddened colors due to the decrease
in metallicity, we adopt the limitations $0<(H-W2)<0.25$,
$-0.1<(Ks-W2)<0.16$, and $-0.04<(Ks-W1)<0.18$ in addition to the limitation
(2) and $Ks<11^{m}$ in order to select the clump
giants. The final sample contains 76 801 and 71 624
clump giants with $b<-50^{\circ}$ and $b>50^{\circ}$, respectively.
It is the stars from this sample that are marked
in Fig. 2 by the larger signs.

It can be seen from Fig. 4b that the change in
$(Ks-W2)$ is a good indicator of the changes in
metallicity. Figures 5c and 5d show the $Ks$ -- $(Ks-W2)$
diagrams for stars in the regions with a radius
of $8^{\circ}$ around the Galactic north (a) and south (b)
poles for $Ks<13^m$. To exclude most of the main
sequence, we introduced the limitation $(Ks-W2)<0.16^{m}$.
The curves indicate the results of our moving
averaging of the data over 55 points. It can be seen
that the mean $\overline{(Ks-W2)}$ changes with distance,
reflecting the change in mean metallicity from $\mathbf Z=0.008\pm0.002$
to $\mathbf Z=0.005\pm0.001$ according to the
isochrones corrected for reddening $E_{(Ks-W2)}\approx0.015^{m}$
as the height $|Z|$ changes from 0.6 to 3 kpc.
It can be seen that the monotonic decrease in metallicity
continues up to $(Ks-W2)\approx0.025^{m}$ at $Ks=12.5^{m}$,
i.e., to $\mathbf Z=0.002$ at $|Z|=6.4$ kpc. All of
these metallicities are determined by this method very
reliably and roughly correspond to the universally accepted
values for these heights.

Next, we considered 18 sky fields: two with a radius
of $8^{\circ}$ around the Galactic poles and the symmetric
sectors in the northern and southern hemispheres
$66^{\circ}<|b|<78^{\circ}$, $-20^{\circ}<l<20^{\circ}$,
$66^{\circ}<|b|<78^{\circ}$, $70^{\circ}<l<110^{\circ}$,
$66^{\circ}<|b|<78^{\circ}$, $160^{\circ}<l<200^{\circ}$,
$66^{\circ}<|b|<78^{\circ}$, $250^{\circ}<l<290^{\circ}$,
$53^{\circ}<|b|<61^{\circ}$, $-15^{\circ}<l<15^{\circ}$,
$53^{\circ}<|b|<61^{\circ}$, $75^{\circ}<l<105^{\circ}$,
$53^{\circ}<|b|<61^{\circ}$, $165^{\circ}<l<195^{\circ}$,
$53^{\circ}<|b|<61^{\circ}$, $255^{\circ}<l<285^{\circ}$.

In each field we performed a moving calculation of
$E_{(H-W1)}/E_{(H-Ks)}$, $E_{(H-W2)}/E_{(H-Ks)}$, $E_{(H-W3)}/E_{(H-Ks)}$,
$E_{(H-W4)}/E_{(H-Ks)}$
(it contains no useful information due to the low accuracy
of the data), and other coefficients as a function
of $r$ with an averaging windows of 200 stars. The
stars are arranged by $r$ and the mean $\overline{r}$, along with
the sought-for coefficients, is calculated for 200 stars
with minimum $r$; the star with minimum $r$ is then
excluded from the set of stars under consideration,
a previously unused star with minimum $r$ is then
introduced instead of it, and the calculations of $\overline{r}$ and
the sought-for coefficients are repeated. As a result,
we obtain several thousand solutions including $\overline{r}$ with
the corresponding set of sought-for coefficients for
each sky field under consideration.

Among all of the analyzed coefficients, below we
consider $E_{(H-W1)}/E_{(H-Ks)}$, $E_{(H-W2)}/E_{(H-Ks)}$, $E_{(H-W3)}/E_{(H-Ks)}$, because
their variations can be directly compared
with the results from Zasowski et al. (2009) and
Gontcharov (2013). This analysis is not an exhaustive
use of the available IR photometry at high
latitudes. Other characteristics of stars and the
medium can also be estimated from the same data.

\section*{RESULTS}

Figure 6 shows the variations of $E_{(H-W2)}/E_{(H-Ks)}$ with heliocentric distance for the region
with a radius of $8^{\circ}$ around the Galactic north (a) and
south (b) poles (black curves with grey error bands).
Smaller values of the coefficient correspond to larger
dust grain sizes and a larger extinction with a longer
wavelength.

The dashes indicate the presumed variation of the
coefficient within 500 pc of the Sun if the dust grain
size distribution in the range $2-11$ microns continues the
trends of the distribution in the range $0.5-2$ microns:
at $r=0$ the dashes indicate $E_{(H-W2)}/E_{(H-Ks)}=1.7$ found by Gontcharov (2013), on average,
for the solar neighborhood, while the variations indicated
by the dashes correspond to the $R_V$ variations
shown in Fig. 1.

First of all, it can be seen from the figure that
the coefficient under consideration outside the disk
(i.e., at $|Z|>0.1$ kpc) is everywhere smaller than
that in the disk. Consequently, as was assumed by
Gontcharov (2013), the fraction of fine dust in the disk
is maximal, while the mean grain size is minimal. It
can be seen that the mean dust grain size outside the
disk decreases with increasing heliocentric distance
and stabilizes or may begin to increase again farther
than 2.4 kpc.

Figure 6c shows the results for the poles together.
It can be seen that the values of $E_{(H-W2)}/E_{(H-Ks)}$
southward and northward of the Sun are very
close at some distances, forming ``nodes'' to which the
distance $r=0$ also refers. These nodes are spaced,
on average, 656 pc apart (the accuracy is discussed
below) and are indicated by the vertical arrows at
the bottom. The variations of the coefficient at the
remaining distances anticorrelate, i.e., the decrease
in coefficient to the south corresponds to its increase
to the north and vice versa. The most prominent
differences between the curves are highlighted in
Fig. 6c as the shaded regions between them. The
results for $R_V$ indicated by the dashes show the
same effect, although this study of $R_V$ has little
in common with that by Gontcharov (2012a): we
considered different classes of stars -- clump giants
instead of RGB and OB stars (no common star),
photometry in the $H$, $Ks$, and $W2$ bands instead of
$B_{T}$, $V_{T}$, $Ks$ (one common band from 2MASS).
Other coefficients, without the $Ks$ band, that are not
shown here, for example, $E_{(H-W2)}/E_{(J-H)}$ and $E_{(H-W2)}/E_{(H-W1)}$,
exhibit the same anticorrelation
to the south and the north. Consequently,
this effect cannot be attributed to the systematic
photometric errors.

To estimate the accuracy of our results, we varied
the averaging window in the range from 100
to 300 stars. In comparison with the window of
200 stars, high-frequency ($\sim10$ pc) variations of the
coefficients under consideration manifest themselves
for the window of 100 stars, which are unlikely
to be real, while the amplitude of the long-period
variations shown in Fig. 6 decreases noticeably for
the window of 300 stars. However, these long-period
variations are observed for any window considered
and, moreover, their phase and period remain within
$\pm20$ pc irrespective of the window size. Thus, it can
be concluded that the nodes are spaced, on average,
$656\pm20$ pc apart. This suggests that the variations
are real, which is quite expectable, given that, as
has been pointed out above, the relative accuracy of
individual $r$ determined by the scatter of individual
$M_{Ks}$ is 15\%. The formal relative accuracy of $\overline{r}$ is then
1\%, i.e., at least 30 pc. Of course, several systematic
effects affecting the result were not taken into account
and cannot be taken into account here. For example,
the sample composition can change with distance
from the Galactic midplane. However, we can hope
for a smooth pattern of these systematic effects depending
on coordinate $Z$. Then, the phase, step, and
constancy of the period may not be determined quite
correctly in the detected variations of the coefficients
under consideration along $Z$, but the very fact of
the existence of periodic or, more precisely, cyclic
variations in the coefficients under consideration has
been established firmly.

We also find a similar anticorrelation between the
variations for the coefficients $E_{(H-W1)}/E_{(H-Ks)}$ and
$E_{(H-W3)}/E_{(H-Ks)}$, which are shown
for the same sky regions in Fig. 7. The presumed
variations of the coefficients within 500 pc of the
Sun based on the results from Gontcharov (2012a)
and $E_{(H-W1)}/E_{(H-Ks)}=1.5$ and $E_{(H-W3)}/E_{(H-Ks)}=2$ from Gontcharov (2013) for
$r=0$ are also indicated here by the dashes. Just as
from $E_{(H-W2)}/E_{(H-Ks)}$, it can be seen from
these coefficients that the dust grains outside the disk
are, on average, larger than those in the disk.

The errors of our results are represented by the
individual vertical bars. The vertical straight lines
indicate the distances marked by the arrows in
Fig. 6. It can be seen that the nodes of variations
for $E_{(H-W1)}/E_{(H-Ks)}$ coincide with those
for $E_{(H-W2)}/E_{(H-Ks)}$, while the analogous
results for $E_{(H-W3)}/E_{(H-Ks)}$ are barely seen
because of the large photometric error in the $W3$
band.

Figure 8 shows the variations of $E_{(H-W2)}/E_{(H-Ks)}$ with heliocentric distance for symmetry
sectors of the sky in the northern (black
curves) and southern (grey curves) hemispheres:
(a) $66^{\circ}<|b|<78^{\circ}$, $-20^{\circ}<l<20^{\circ}$,
(b) $66^{\circ}<|b|<78^{\circ}$, $70^{\circ}<l<110^{\circ}$,
(c) $66^{\circ}<|b|<78^{\circ}$, $160^{\circ}<l<200^{\circ}$,
(d) $66^{\circ}<|b|<78^{\circ}$, $250^{\circ}<l<290^{\circ}$,
(e) $53^{\circ}<|b|<61^{\circ}$, $-15^{\circ}<l<15^{\circ}$,
(f) $53^{\circ}<|b|<61^{\circ}$, $75^{\circ}<l<105^{\circ}$,
(g) $53^{\circ}<|b|<61^{\circ}$, $165^{\circ}<l<195^{\circ}$,
(h) $53^{\circ}<|b|<61^{\circ}$, $255^{\circ}<l<285^{\circ}$.
The same
effects as those for the poles are seen: the results
for the hemispheres are similar, anticorrelate between
themselves, the dust outside the disk is everywhere
coarser than that in the disk. However, these effects
are less pronounced than those for the poles due to the
influence of the errors in the photometric distances
derived from Eq. (1). The maxima and minima of
the coefficient variations in Fig. 8 are shifted along $r$
relative to those in Fig. 6 approximately with the
coefficient $(\sin b)^{-1}$. Consequently, the coarse and
fine dust at high latitudes is located in layers that
alternate along $|Z|$ and are approximately parallel to
the Galactic plane. This also explains the variations of
$E_{(H-W2)}/E_{(H-Ks)}$ with latitude visible in the
figure: at $53^{\circ}<|b|<61^{\circ}$ it is, on average, smaller
than that at $66^{\circ}<|b|<78^{\circ}$. It can also be seen from
the figure that there are no variations of the coefficient
with longitude everywhere at high latitudes. Therefore,
although here we considered only individual
fields of the celestial sphere, our results undoubtedly
refer to all high latitudes. However, as the size of
the fields increases, the detected variations of the
coefficients under consideration become invisible due
to the errors of the photometric distances.

The detected consistency of the variations in the
characteristics of the interstellar medium southward
and northward of the Sun within at $-2.5<Z<2.5$ kpc, i.e., 5 kpc, is an unexpected result
that requires a further confirmation. It is important
that it could not be obtained before the appearance
of WISE results, the first (in the history of astronomy)
accurate IR photometry at high latitudes. This
result suggests the connections between widely separated
regions of the seemingly very tenuous interstellar
medium, a unified large-scale structure of the
medium within at least 5 kpc. This structure may
have emerged long ago during the formation of our
Galaxy and has been maintained for billions of years.

It may worth considering this structure as an alternation
of layers of not fine and coarse dust but
sorted and unsorted dust. The cause of the sorting
is apparently the same as that in the spiral pattern
of the disk—the density wave and the corresponding
periodic star formation. Based on the theory and observations,
Fridman and Khoperskov (2011) showed
that the motions of the Galactic medium along $Z$
together with the motions along the Galactic plane
constitute a unified three-component velocity vector
field in the spiral density wave. It then worth noting
that the structure we found here has a spatial step
of $656\times2=1312$ pc, a value close to the typical
distance between the Galactic spiral arms.

The anticorrelation in the arrangement of dust
layers with different properties in the southern and
northern hemispheres may be considered as a possible
result of the periodic alternation of star formation
due to a sharp increase in the density of the
protogalactic medium alternately in the southern and
northern hemispheres. This, in turn, can be explained
by the tidal effect from a protogalaxy's high-latitude
satellite.

\section*{CONCLUSIONS}

This study showed the possibility of jointly using
accurate near- and mid-IR photometry from the
2MASS and WISE catalogues, respectively, to analyze
the variations of the extinction law and the corresponding
interstellar dust properties at high Galactic
latitudes.

Owing to the high accuracy of the data, the large
number of stars considered, and the convenience of
using multicolor IR photometry, the modified method
of extrapolation of the extinction law applied to clump
giants turned out to be efficient for separating the
spatial variations of the sample composition, metallicity,
reddening, and properties of the medium. As
a result, we found effects whose detection was not
possible before the appearance of the 2MASS and
WISE catalogues.

We analyzed the variations of the coefficients
$E_{(H-W1)}/E_{(H-Ks)}$, $E_{(H-W2)}/E_{(H-Ks)}$ and $E_{(H-W3)}/E_{(H-Ks)}$ in 18 fields at high
latitudes ($|b|>50^{\circ}$). The dependence of our results
on Galactic latitude and longitude is insignificant,
which allows the results to be assigned to high
latitudes as a whole.

As a result, we showed that the ratio of shortwavelength
extinction to long-wavelength one everywhere
outside the Galactic disk (i.e., at $|Z|>100$ pc)
is smaller than that in the disk and, accordingly, the
mean dust grain size is larger, while the grain size
distribution is shifted toward coarse dust. Specifically,
the mean grain size initially increases sharply
with $|Z|$, then decreases gradually, approaching a
value typical of the disk at $|Z|\approx2.4$ kpc, and, further
out, stabilizes or may increase again.

The coefficients under consideration change with
coordinate $Z$ with a period of about $1312\pm40$ pc,
coinciding every $656\pm20$ pc to the south and the
north and showing a significant anticorrelation between
their values in the southern and northern hemispheres
at intermediate $Z$. A unified large-scale
periodic structure of the interstellar medium at high
latitudes within at least 5 kpc manifests itself in this
way. These periodic variations were also found in
the three-dimensional map of variations in extinction
coefficient $R_V$ constructed by Gontcharov (2012a)
through the reduction of different photometric data for
stars of different classes.

The detected structure of the medium outside the
Galactic disk is assumed to be maintained by a density
wave just as the spiral structure of the disk.

\section*{ACKNOWLEDGMENTS}

In this study, we used results from the Two Micron
All Sky Survey (2MASS) and Wide-field Infrared
Survey Explorer (WISE) as well as resources from
the Strasbourg Astronomical Data Center (Centre de
Donn\'ees astronomiques de Strasbourg). This study
was supported by Program P21 of the Presidium of
the Russian Academy of Sciences and the Ministry
of Education and Science of the Russian Federation
under contract 8417.

\newpage

\begin{figure}
\includegraphics{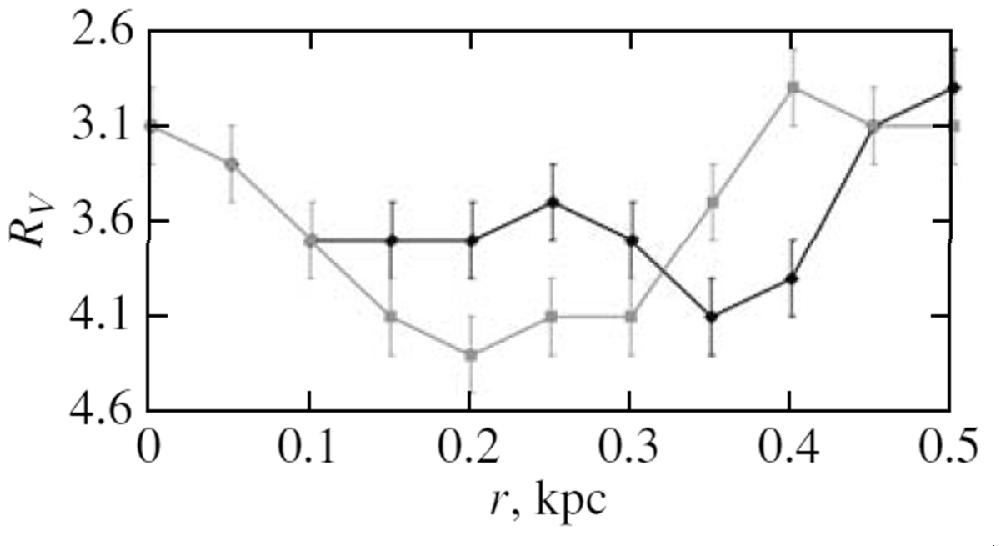}
\caption{Coefficient $R_V$ versus heliocentric distance for
the region with a radius of $8^{\circ}$ around the Galactic north
(black diamonds, curve, and vertical error bars) and south
(grey squares, curve, and vertical error bars) poles.}
\label{paper4}
\end{figure}

\begin{figure}
\includegraphics{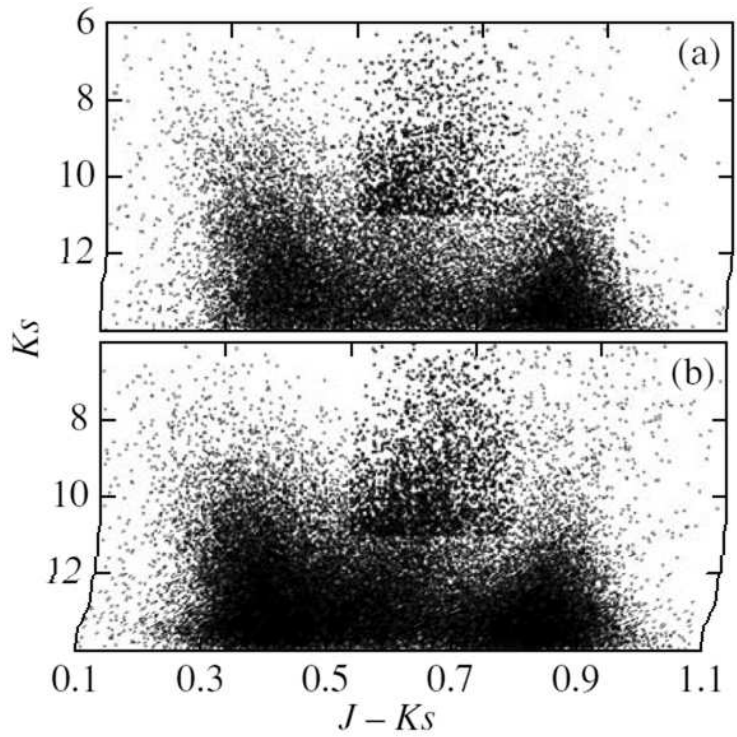}
\caption{Distribution of the stars under consideration on
the $(J-Ks)$ -- $Ks$ diagram (a) in the sky region with a
radius of $8^{\circ}$ around the Galactic north pole and (b) in
the sky sector $-15^{\circ}<l<15^{\circ}$, $-61^{\circ}<b<-53^{\circ}$. The
selected stars are marked by the larger signs in the upper
parts of the graphs.}
\label{jkk}
\end{figure}

\begin{figure}
\includegraphics{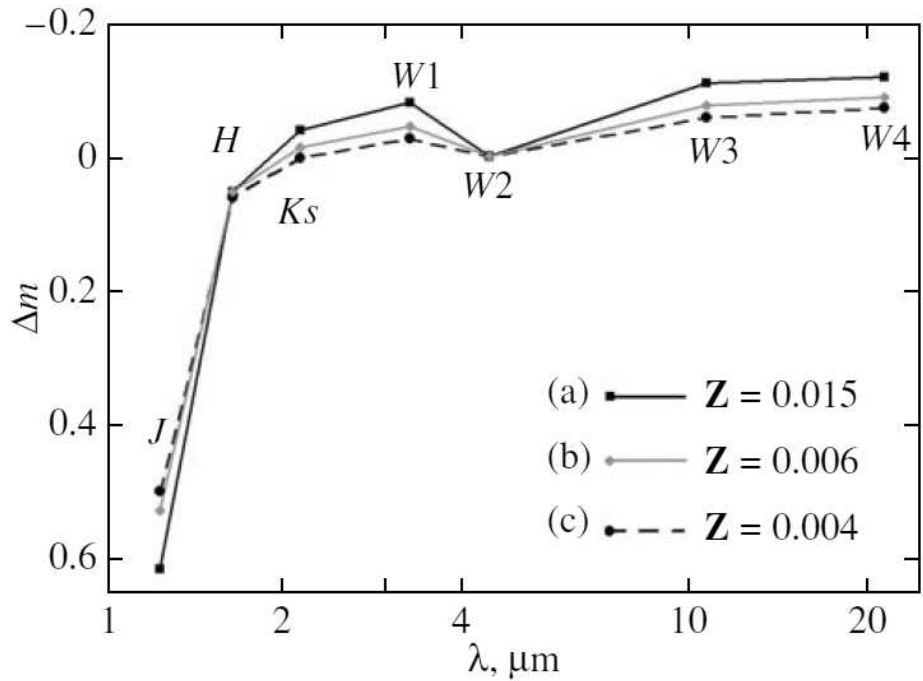}
\caption{Theoretical spectral energy distribution for a typical
clump giant with an age of $10\times10^9$ yr for metallicities
(a) $\mathbf Z=0.015$ (black squares and solid curve), (b) $\mathbf Z=0.006$
(grey diamonds and solid curve), and (c) $\mathbf Z=0.004$
(black circles and dashed line) in the (from left to right)
$J$, $H$, $Ks$, $W1$, $W2$, $W3$, and $W4$ bands expressed in
magnitudes relative to the $W2$ magnitude.}
\label{sedfeh}
\end{figure}

\begin{figure}
\includegraphics{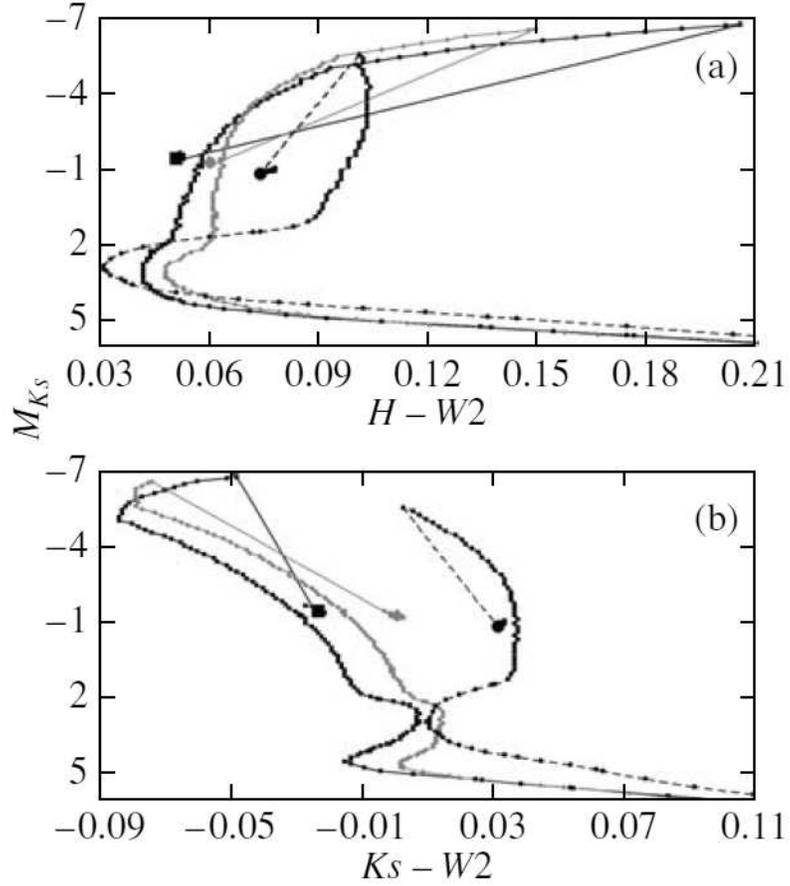}
\caption{Theoretical isochrones before the stage of a
clump giant (highlighted by the large symbols) for
stars with metallicities $\mathbf Z=0.008$ (black squares and
solid curve), $\mathbf Z=0.004$ (grey diamonds and solid
curve, and $\mathbf Z=0.0004$ (black circles and dashed
line) on the (a) $(H-W2)$ -- $M_{Ks}$ and (b) $(Ks-W2)$ -- $M_{Ks}$
diagrams according to the Padova database
((http://stev.oapd.inaf.it/cmd; Marigo et al. 2008; Bressan et al. 2012).}
\label{izo}
\end{figure}

\begin{figure}
\includegraphics{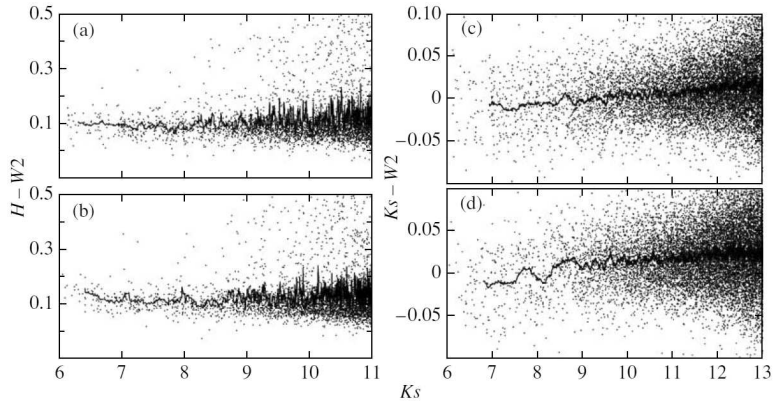}
\caption{$Ks$ -- $(H-W2)$ and $Ks$ -- $(Ks-W2)$ diagrams for stars in the regions with a radius of $8^{\circ}$
around the Galactic north ((a)
and (c)) and south ((b) and (d)) poles. Only the stars with $(Ks-W2)<0.16^{m}$ were left on the $Ks$ -- $(Ks-W2)$ diagram.
The curves indicate the results of our moving averaging of the data over 15 and 55 points for $(H-W2)$ and $(Ks-W2)$,
respectively.}
\label{hw2kw2}
\end{figure}

\begin{figure}
\includegraphics{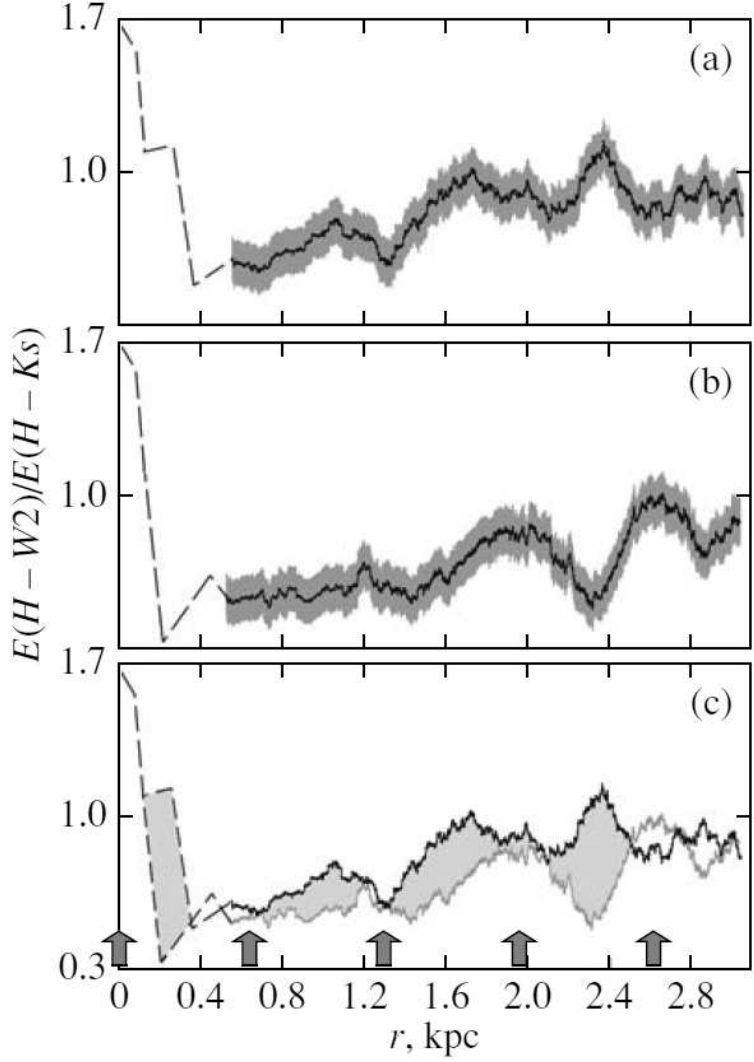}
\caption{$E_{(H-W2)}/E_{(H-Ks)}$ versus heliocentric distance
for the region with a radius of $8^{\circ}$ around the Galactic
north (a) and south (b) poles (black curves with grey
error bands). The dashes indicate the presumed variation
of the coefficient within 500 pc of the Sun; (c) combined
results for the poles: the most prominent differences between
the curves are highlighted as the shaded regions
between the plots; the arrows indicate the distances at
which the results for the poles are similar.}
\label{ngpsgp2}
\end{figure}

\begin{figure}
\includegraphics{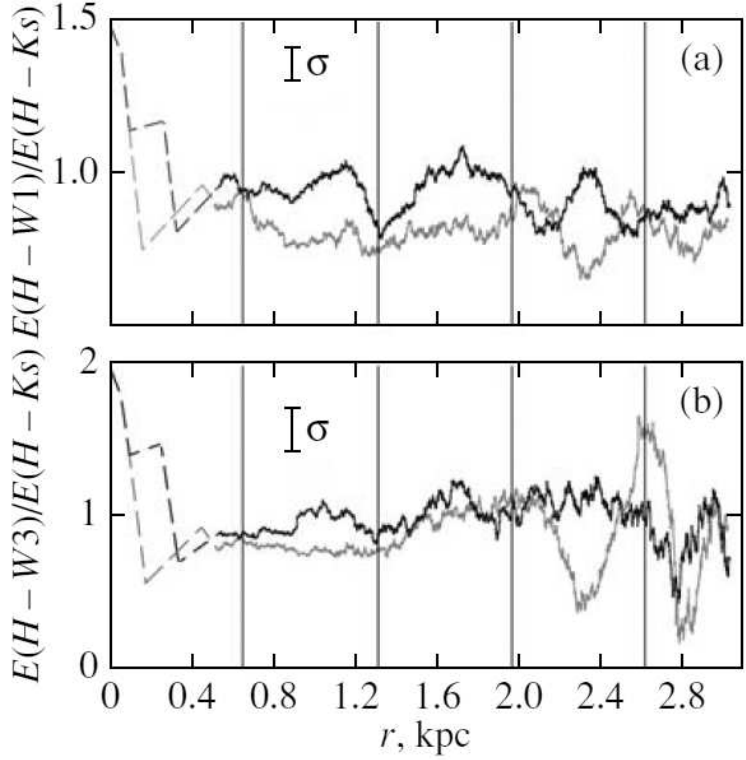}
\caption{$E_{(H-W1)}/E_{(H-Ks)}$ (a) and $E_{(H-W3)}/E_{(H-Ks)}$ (b) versus heliocentric distance for the
region with a radius of $8^{\circ}$ around the Galactic north
(black curves) and south (grey curves) poles. The dashes
indicate the presumed variations of the coefficients within
500 pc of the Sun. The errors are represented by the
individual vertical bars. The vertical straight lines indicate
the distances marked by the arrows in Fig. 6.}
\label{ngpsgp134}
\end{figure}

\begin{figure}
\includegraphics{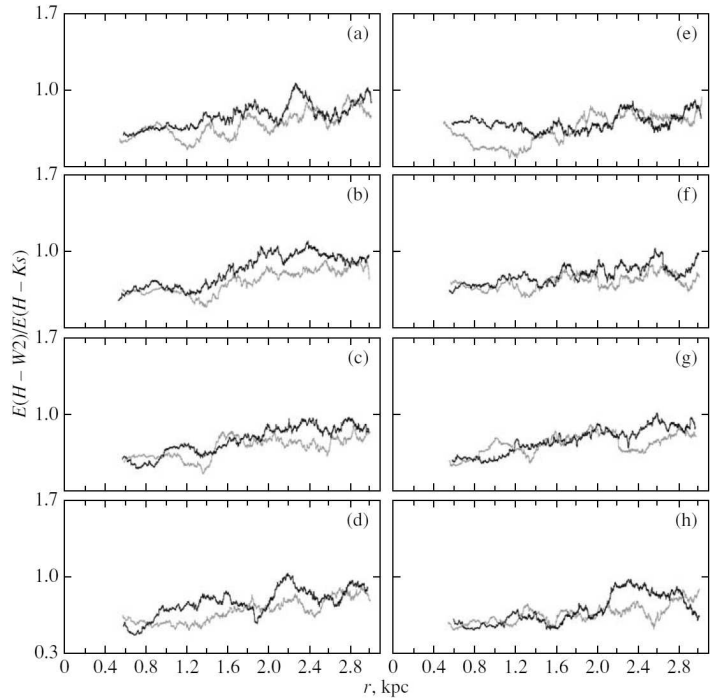}
\caption{$E_{(H-W2)}/E_{(H-Ks)}$ versus heliocentric distance for symmetric sectors of the sky in the northern (black
curves) and southern (grey curves) hemispheres:
(a) $66^{\circ}<|b|<78^{\circ}$, $-20^{\circ}<l<20^{\circ}$,
(b) $66^{\circ}<|b|<78^{\circ}$, $70^{\circ}<l<110^{\circ}$,
(c) $66^{\circ}<|b|<78^{\circ}$, $160^{\circ}<l<200^{\circ}$,
(d) $66^{\circ}<|b|<78^{\circ}$, $250^{\circ}<l<290^{\circ}$,
(e) $53^{\circ}<|b|<61^{\circ}$, $-15^{\circ}<l<15^{\circ}$,
(f) $53^{\circ}<|b|<61^{\circ}$, $75^{\circ}<l<105^{\circ}$,
(g) $53^{\circ}<|b|<61^{\circ}$, $165^{\circ}<l<195^{\circ}$,
(h) $53^{\circ}<|b|<61^{\circ}$, $255^{\circ}<l<285^{\circ}$.}
\label{bl}
\end{figure}

\end{document}